\documentclass[groupedaddress,
 reprint,
 nofootinbib,
 amsmath,amssymb,
 aps,
 prd
]{revtex4-1}

\usepackage{graphicx}
\usepackage{dcolumn}
\usepackage{bm}
\usepackage[hyperindex,breaklinks]{hyperref}
\usepackage[utf8]{inputenc}
\usepackage{esint}
\usepackage[capitalise]{cleveref}

\usepackage{aas_macros}

\usepackage{comment}
\usepackage{color}

\begin{document}

\title{Tidal disruption of fuzzy dark matter subhalo cores}

\author{Xiaolong Du}
\email{xiaolong@astro.physik.uni-goettingen.de}
\author{Bodo Schwabe}
\author{Jens C. Niemeyer}
\author{David B\"{u}rger}
\affiliation{%
Institut f\"ur Astrophysik, Georg-August-Universit\"at G\"ottingen, Friedrich-Hund-Platz 1, D-37075 G\"ottingen, Germany
}%

\date{\today}

\begin{abstract}
We study tidal stripping of fuzzy dark matter (FDM) subhalo cores using simulations of the Schr\"{o}dinger-Poisson equations and analyze the dynamics of tidal disruption, highlighting the differences with standard cold dark matter. Mass loss outside of the tidal radius forces the core to relax into a less compact configuration, lowering the tidal radius. As the characteristic radius of a solitonic core scales inversely with its mass, tidal stripping results in a runaway effect and rapid tidal disruption of the core once its central density drops below $4.5$ times the average density of the host within the orbital radius.
Additionally, we find that the core is deformed into a tidally locked ellipsoid with increasing eccentricities until it is completely disrupted. Using the core mass loss rate, we compute the minimum mass of cores that can survive several orbits for different FDM particle masses and compare it with observed masses of satellite galaxies in the Milky Way.

\end{abstract}


\maketitle

\section{Introduction}\label{sec:intro}

Ultralight bosonic fields with masses $m \gtrsim 10^{-22}$ eV are viable dark matter candidates [fuzzy dark matter (FDM)] with interesting phenomenology on galactic scales even with purely gravitational interactions \cite{Sin1994,Hu:2000ke}. They arise naturally as axionlike particles (ALPs) in theories with weakly broken shift symmetries and are a common biproduct of string theory compactifications \cite{Witten:1984dg,Svrcek:2006yi,Arvanitaki:2009fg}. For a recent review, see \cite{Marsh2015}.

At present, the strongest bounds on the minimum mass of FDM particles are derived from the (linear) suppression of small-scale power by scalar field gradients (the so-called ``quantum Jeans scale''). For example, current observations of the Lyman-alpha forest have been interpreted as implying $m > 2\times10^{-21}$eV \cite{Irsic:2017yje,Kobayashi:2017jcf} but systematic uncertainties are still disputed \cite{Zhang:2017chj}. Slightly weaker constraints follow from the cosmic microwave background and large scale structure~\cite{Hlozek:2014lca,Hlozek:2016lzm,Hlozek:2017zzf,Nguyen:2017zqu}, the abundance of dark matter halos and subhalos~\cite{Marsh2014,Du:2016zcv,Menci:2017nsr,Schive:2017biq}, or high-$z$ galaxy luminosity functions and  reionization~\cite{Bozek:2014uqa,Schive2016,Corasaniti:2016epp}. 

In the nonlinear, nonrelativistic regime of gravitational collapse described by the Schr\"{o}dinger-Poisson (SP) equations, FDM differs from standard cold dark matter (CDM) by the formation of solitonic halo cores whose density profiles are similar to boson star solutions of the SP equations \cite{Schive2014a,Schive2014b,Schwabe:2016rze,Veltmaat:2016rxo}.  If $m \sim 10^{-22}$ eV, these cores can potentially account for the observed rotation curves of dwarf galaxies \cite{Schive2014a,Marsh2015b,Calabrese2016,Chen:2016unw,Gonzales-Morales:2016mkl}. 

FDM halo substructure is shaped both by the linear suppression of small-scale perturbations and the nonlinear formation of solitonic cores. The former gives rise to a mass-dependent cutoff of the subhalo mass function (SHMF), while the latter governs the density profiles of subhalos and their stability against tidal disruption \cite{Marsh2014,Du:2016zcv}. Information about halo substructure in the Milky Way or other galaxies, gathered, e.g., from gravitational lensing of subhalos \cite{Metcalf:2001ap,Bradac:2001mv,Dalal:2001fq,Keeton:2008gq,Vegetti:2009,Vegetti:2014,Hezaveh2014,Hezaveh2016}, the subhalo abundance \cite{Vale:2004yt,Conroy:2005aq}, or tidal streams \cite{Carlberg:2009,Sanders2016,Erkal:2016} thus has the potential to place stringent constraints on FDM scenarios.

For CDM, the substructure of dark matter halos has been studied intensively both with N-body simulations (e.g.~\cite{Diemand:2008in,Springel:2008cc}) and semianalytic models (SAMs) (e.g.~\cite{Penarrubia:2004et,vandenBosch:2004zs,Zentner:2004dq,Kampakoglou:2006uf,Jiang:2014nsa,Pullen:2014gna}). For FDM, on the other hand, cosmological simulations of the SP equations with the required resolution are still unavailable. \cite{Du:2016zcv} used a modified version of the semianalytic code \textsc{Galacticus} \cite{Benson2012,Benson:2012su} to study the effects of tidal stripping and dynamical friction on FDM subhalos. In the absence of simulations to calibrate the model, simple approximations had to be made which resulted in significant effects on the predicted SHMF. Under the assumption that the incoherent matter in FDM subhalos with an NFW-like density profile behaves similarly to CDM, their solitonic cores will be stripped clean after a certain number of orbits and exposed to tidal forces. In one particular case, analyzed in \cite{Du:2016zcv}, cores were assumed to be stable against tidal stripping, giving rise to a pronounced peak in the SHMF. The sensitivity of the SHMF to different prescriptions for tidal stripping clearly motivates a detailed study of FDM cores under tidal stress. 

In \cite{Hui:2016ltb},  tidal mass-loss of a solitonic core orbiting inside a host halo is computed in a ``tunneling approximation'' by adding a spherical tidal potential to the time-independent SP equations. The mass loss rate is obtained from the imaginary part of the (complex) energy eigenvalue $E$. Since both tidal and gravitational potentials are taken to be time independent, so is the tidal radius. Consequently, the mass loss is fully characterized by the decreasing amplitude of the wave function.

As shown below, the actual dynamics are more involved. Once mass outside the tidal radius is removed, the core relaxes to a new ground state with smaller mass and accordingly larger core radius. In the process, mass is transferred through the tidal radius and subsequently stripped away resulting in a continuous mass loss.

This process has a classical analog. After the outer parts of the satellite are stripped away, the remnant is no longer in virial equilibrium and needs to convert kinetic to potential energy in order to re-equilibrate \cite{vandenBosch:2017ynq,vandenBosch:2018tyt}. The resulting configuration has a larger characteristic radius and decreased density, achieved by an outwards directed mass transfer through the tidal radius which, in turn, shrinks as a result of the lowered enclosed mass. However, for CDM, as shown in \cite{vandenBosch:2018tyt}, this process is usually not sufficient to disrupt the subhalo.

In this work, we present numerical simulations that quantify the mass loss rate from tidal stripping and estimate the survival time of satellite galaxies in the Milky Way. We also investigate the shape of the solitonic core and find that it does not relax into a spherically symmetric ground state but rather into a Riemann-S ellipsoid as analyzed in \cite{RindlerDaller:2011kx}. It becomes tidally locked with increasing ellipticity until being completely disrupted.

The paper is organized as follows. In Sec. \ref{sec:classical_quantum}, we repeat and extend the analytic arguments of \cite{Hui:2016ltb}. We then outline the implementation of our numerical simulations in Sec. \ref{sec:sim}. We summarize our results in Secs. \ref{sec:results}--\ref{sec:sat_galaxy} and conclude in Sec. \ref{sec:conclusions}.

\section{Classical and tunneling tidal radius}
\label{sec:classical_quantum}

A satellite halo orbiting the host halo loses its mass due to the tidal force of the host halo, i.e. the tidal stripping effect. Considering a satellite orbiting its host with synchronous rotation, i.e. the angular velocity of self-rotation equals the orbital angular velocity, the tidal radius can be calculated from classical Newtonian dynamics \cite{King:1962}:
\begin{equation}
r_t=\left(\frac{G M_{\rm sat}(<r_t)}{\omega^2-d^2\Phi/dx^2}\right)^{1/3}\,,
\label{eq:x_t_c}
\end{equation}
where $M_{\rm sat}$ is the satellite mass enclosed within the tidal radius, $\omega$ is the angular velocity of the satellite, $\Phi$ is the gravitational potential of the host halo, and $x$ is the distance to the host's center. Assuming a circular orbit of the satellite and most of the host mass to be within the orbital radius, we have
\begin{equation}
\frac{d^2\Phi}{d x^2}=-\frac{2GM_{\rm host}(<x_{\rm sat})}{x_{\rm sat}^3}=-2\omega^2\,.
\label{eq:dPhi}
\end{equation}
Then the tidal radius can be written as
\begin{equation}
r_t=\left(\frac{GM_{\rm sat}(<r_t)}{3\omega^2}\right)^{1/3}\,.
\label{eq:r_t_c2}
\end{equation}

In \cite{Hui:2016ltb}, tidal stripping of FDM halos is treated quantum-mechanically by adding a spherical tidal potential to the Schr\"{o}dinger equation. The authors propose that mass inside the tidal radius can be stripped in sufficiently long time due to tunneling.

Following this approach, we first consider a simple system in which the solitonic core is subject to a spherically symmetric tidal potential $\Phi_t=-\gamma\omega^2 r^2$ (here, $r$ is the distance to the center of the satellite). Note that \cite{Hui:2016ltb} use $\gamma=\frac{3}{2}$ which includes the effect of the centrifugal force owing to synchronous rotation of the satellite, assuming it to be a rigid body. However, a solitonic core forms an irrotational Riemann-S ellipsoid when subject to the tidal force as discussed in Sec. \ref{sec:spinning}. Therefore, for a solitonic core, $\gamma$ in the tidal potential should be between $1$ (without self-rotation) and $\frac{3}{2}$ (with uniform self-rotation that equals the orbital angular velocity). To be comparable with \cite{Hui:2016ltb} we fix $\gamma$ to $\frac{3}{2}$ unless specified otherwise. 

Working in a coordinate system centered on the satellite, the SP equations become
\begin{align}
i \hbar \frac{\partial \psi}{\partial t}
&= -\frac{\hbar^2}{2m_a}\nabla^2\psi+m_a\left(\Phi-\frac{3}{2}\omega^2 r^2\right)\psi\,,
\label{eq:SP1}\\
\nabla^2\Phi&= 4\pi G m_a |\psi|^2\,.
\label{eq:SP2}
\end{align}

In \cite{Hui:2016ltb}, the authors decompose the wave function $\psi(r,t)=\phi(r)\exp(-i E t)$ to get the time-independent SP equations. Assuming the energy eigenvalue $E$ to be complex, they obtain the mass loss rate from the imaginary part of $E$,
\begin{equation}
\frac{\dot{M}}{M}=\frac{\dot{\rho}}{\rho}=2\,{\rm Im}(E)\,,
\label{eq:m_loss}
\end{equation}
which only depends on the density ratio between the central density of the soliton $\rho_c$ and the average density of the host within the orbital radius $\overline{\rho}_{\rm host}$, i.e.  $\mu\equiv\rho_c/\overline{\rho}_{\rm host}$. By solving the eigenvalue problem as in \cite{Hui:2016ltb}, we find a fitting formula for the imaginary part of $E$:
\begin{equation}
{\rm Im}(E)= -T_{\rm orbit}^{-1}\exp\left[a \left(\frac{3}{2\gamma}\mu\right)^2+b \left(\frac{3}{2\gamma}\mu\right)+c\right]\,,
\label{eq:m_loss_fitting}
\end{equation}
with the best-fitting parameters $\{a,b,c\}=\{5.89794\times10^{-5},-8.72733\times10^{-2},1.6774\}$. Here, $T_{\rm orbit}\equiv {2\pi}/{\omega}$ is the orbital period.

However, as the gravitational potential depends on the amplitude of $\psi(r,t)$, in principle, we cannot separate $\psi(r,t)$ into two parts which purely depend on time and radius, respectively. Therefore, the treatment in \cite{Hui:2016ltb} can only be seen as an approximation for small enough time scales on which the gravitational potential can be treated as time independent. As the solitonic core loses some of its mass and becomes less dense, it is increasingly vulnerable to tidal forces.

\begin{figure}[htbp]
\includegraphics[width=\columnwidth]{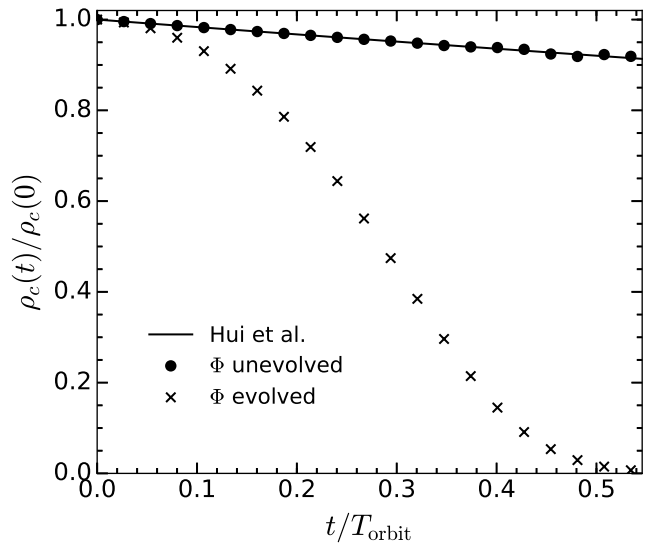}
\caption{Evolution of the core's central density with unevolved (dots) and evolved (crosses) gravitational potential compared to the prediction in \cite{Hui:2016ltb}.}
\label{fig:rho_c_t_Phi}
\end{figure}

To test these arguments, we set up two special simulations using a pseudospectral solver (see the next section for details). First, we assume that the gravitational potential $\Phi$ in \cref{eq:SP1,eq:SP2} does not change with time. By solving the eigenvalue problem, we obtain the ground eigenstate with the parameter $\mu=50$ and use it as the initial condition. Then we solve the time-dependent Schr\"{o}dinger equation [\cref{eq:SP1}] assuming that $\Phi$ does not change. Finally, for comparison we allow $\Phi$ to be time dependent and solve the full SP equations with the same initial condition. The results are shown in \cref{fig:rho_c_t_Phi}. As can be seen, if $\Phi$ does not change with time, the evolution of the core's central density exactly matches the prediction in \cite{Hui:2016ltb}. But if we consider the full nonlinear problem, the evolution of the central density is consistent with the prediction only at the very beginning, afterwards the central density decreases more quickly.

The density profile of solitonic cores can be approximated by
\begin{equation}
\rho_s(r) = \frac{\rho_c}{[1+(\alpha r/r_c)^2]^8}\,,
\label{eq:rho_c}
\end{equation}
where $\rho_c$ is the central density and $r_c$ is the radius where the density drops to half of the central density \cite{Schive2014a,Schive2014b,Marsh2015b}. We will call $r_c$ the core radius and set  $\alpha = 0.302$ as in \cite{Schive2014a,Schive2014b} hereafter. As a result of the scaling relation $(r,\psi,\Phi,E,\omega)\rightarrow(r/\lambda,\lambda^2\psi,\lambda^2\Phi,\lambda^2 E,\lambda^2\omega)$, it follows that $\rho_c\sim r_c^{-4}$. Calculating the gravitational potential of a solitonic core from \cref{eq:rho_c}, the tidal radius can be obtained via \cref{eq:r_t_c2}. It is easy to check with the help of the scaling relation that the tidal radius in units of the core radius, $r_t/r_c$, only depends on the density ratio $\mu$.

\begin{figure}[htbp]
\includegraphics[width=\columnwidth]{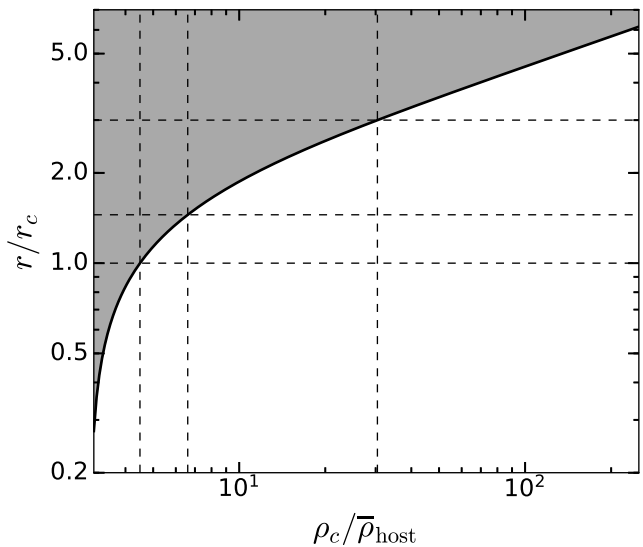}
\caption{Regions inside (blank region) and outside (shaded region) the tidal radius with respect to the density ratio $\mu\equiv\rho_c/\overline{\rho}_{\rm host}$. The solid line shows the tidal radius computed from \cref{eq:r_t_c2}. The horizontal lines mark the radii enclosing $95\%$ (top), $50\%$ (middle) and $25\%$ (bottom) of the total soliton mass, respectively. When $\mu<4.5$, the tidal radius is smaller than the core radius. Thus, the solitonic core becomes unstable and is quickly disrupted. 
}
\label{fig:pic_r_t}
\end{figure}

\Cref{fig:pic_r_t} shows regions inside (blank region) and outside (shaded region) the tidal radius with respect to the density ratio. For $\mu\gtrsim30.4$, more than $95\%$ of the total soliton mass is within the tidal radius.

\section{Numerical methods}\label{sec:sim}

To investigate fully time-dependent tidal stripping of solitonic subhalo cores with increasingly relaxed symmetry assumptions, we conducted numerical simulations of the spherically symmetric [\cref{eq:SP1,eq:SP2}] and the full three-dimensional problem.

In the spherically symmetric case, we use the ground state of \cref{eq:SP1,eq:SP2} as initial conditions
and work with dimensionless quantities as in \cite{Guzman2004}. All the results presented here can be rescaled to restore the physical units. The boundary is set at $r_{\rm max}=280$ (as a reference, the initial core radius is at about $r_c=1.308$). We implement an absorbing boundary condition by adding a ``sponge" (imaginary potential) in the outer regions of the grid,
\begin{equation}
V(r)=-\frac{i}{2}V_0\{2+\tanh{[(r-r_s)/\delta]}-\tanh{(r_s/\delta)}\}\Theta(r-r_p) \,,
\label{eq:boundary_con}
\end{equation}
where $\Theta$ is the Heaviside function \cite{Guzman2004}. We set $r_p=2/7 \,r_{\rm max}$, $r_s=(r_{\rm max}+r_p)/2$, $\delta=(r_{\rm max}-r_p)$, and  $V_0=2$.

In the full three-dimensional case, we work in coordinates centered on the host.
The host is treated as a uniform sphere with mass $M_{\rm host}$ and a radius of roughly $10$ times the cell size. Contrary to the spherically symmetric case, no eigenstate is available. We therefore use soliton solutions as initial conditions and assume periodic boundary conditions. The soliton is placed initially at a distance of $D=25$ to the host and is given an initial velocity $v_0=(G M_{\rm host}/D)^{1/2}$. The simulated box has a length of $160$ on each side and totally $480^3$ cells so that the core radius is covered by at least $4$ cells. We have verified that the artificial ``sponge” is not necessary in this case over the entire simulation time.

To solve the SP equations, we have developed a fourth-order pseudospectral solver. It provides fourth-order convergence in time and spectral convergence in space. Compared to previous second-order pseudospectral methods, e.g. \cite{Woo:2008nn,Mocz:2017wlg}, our code is about $6$ times faster in getting comparable accuracy.

The wave function is advanced in time by a unitary transformation,
\begin{equation}
\psi(t+\Delta t)=\exp(-i H \Delta t)\psi(t),
\label{eq:psi_dt}
\end{equation}
where $H$ is the Hamiltonian of the system which can be split into the kinetic part $K$ and the potential part $W$, i.e. $H=K+W$. In general, the operator $\exp(-i H \Delta t)$ can be expanded as
\begin{equation}
\exp(-i H \Delta t)=\prod_{i}\exp(-i t_i K \Delta t)\exp(-i v_i W \Delta t),
\label{eq:exp_H}
\end{equation}
where $t_i$ and $v_i$ are parameters to be determined by the requirements of the chosen order. For example, to second order we obtain the well-known leapfrog method,
 \begin{equation}
\exp(-i H \Delta t)=e^{-\frac{i}{2} W \Delta t}e^{-i K \Delta t}e^{-\frac{i}{2} W \Delta t}+O(\Delta t^3),
\label{eq:exp_H_O2}
\end{equation}
which is also referred to as the ``kick-drift-kick" formulation. If we exchange the operators $K$ and $W$ in \cref{eq:exp_H_O2}, i.e. update the position first, we arrive at the ``drift-kick-drift" formulation which also has second-order accuracy.

In our simulations, we implement the fourth-order algorithm proposed by McLachlan \cite{McLachlan:1995},
\begin{eqnarray}
e^{-i H \Delta t}\approx&&e^{-i v_2 W\Delta t}e^{-i t_2 K\Delta t}e^{-i v_1 W\Delta t}e^{-i t_1 K\Delta t} e^{-i v_0 W\Delta t}\nonumber\\
&&e^{-i t_1 K \Delta t}e^{-i v_1 W \Delta t}e^{-i t_2 K\Delta t}e^{-i v_2 W\Delta t}\,,
\label{eq:exp_H_M}
\end{eqnarray}
where
\begin{eqnarray}
&&v_1=\frac{121}{3924}(12-\sqrt{471}),\quad
w=\sqrt{3-12 v_1+9 v_1^2},\nonumber\\
&&t_2=\frac{1}{4}\left(1-\sqrt{\frac{9 v_1-4+2 w}{3 v_1}}\right)\,,\quad
t_1=\frac{1}{2}-t_2,\nonumber\\
&&v_2=\frac{1}{6}-4 v_1 t_1^2,\quad
v_0=1-2(v_1+v_2).
\label{eq:M_par}
\end{eqnarray}
Compared to the leapfrog method, it is much more accurate. Note that the kinetic operator is performed in Fourier space, while the potential operator is performed in real space. In Fourier space the kinetic operator can be computed in a very simple way: $e^{-i K}\hat{\psi}=e^{-i \hbar^2 k^2/2/m_a}\hat{\psi}$. The potential $W$ is obtained by solving the Poisson equation via a spectral method \cite{Woo:2008nn,Mocz:2017wlg}. We have verified that our code has fourth-order convergence in time by simulating mergers of multiple solitons (see \cref{sec:appendix}).

\section{Simulations of tidal stripping}
\label{sec:results}

\subsection{Spherically symmetric approximation}\label{sec:1D}

We simulate the evolution of cores with different initial density ratios $\mu$. The mass within the tidal radius $M_t$ is computed at different times. The mass loss rate is then calculated by
\begin{align}
\frac{\dot{M}_t}{M_t} &= \frac{1}{M_t}\frac{d}{d t}\int_0^{r_t(t)} 4\pi r^2 \rho(r,t) dr \nonumber\\
&= \frac{1}{M_t}\int_0^{r_t(t)} 4\pi r^2 \partial_t \rho(r,t) dr+\frac{1}{M_t}4\pi r_t^2\rho(r_t,t)\dot{r}_t\,,
\label{eq:m_loss_2}
\end{align}
where the first term can be interpreted as mass transfer through the tidal radius and the second term corresponds to effects of a decreasing tidal radius. Using Gauss' theorem we can rewrite the first term as
\begin{equation}
\frac{1}{M_t}\int_0^{r_t(t)} 4\pi r^2 \partial_t \rho(r,t) dr=-\frac{1}{M_t}\oiint_{r=r_t}\rho(r,t)\mathbf{v}(r,t)\cdot d\mathbf{S}\,,
\label{eq:mass_loss_qt}
\end{equation}
where $\mathbf{v}$ is the velocity field.

\begin{figure}[htbp]
\includegraphics[width=0.95\columnwidth]{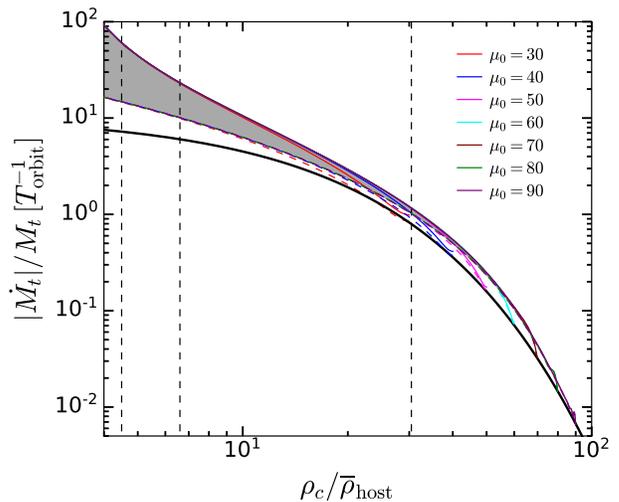}
\caption{Mass loss rate for different initial density ratios. A spherical tidal field is assumed. The solid black line shows the prediction in \cite{Hui:2016ltb} using the fitting formula \cref{eq:m_loss_fitting}. The vertical dashed lines are the same as in \cref{fig:pic_r_t}. The solid lines with different colors give the total mass loss rate. The dashed colored lines show the mass loss rate by mass transfer through the tidal radius, while the shaded region corresponds to the mass loss rate from the decreasing tidal radius.}
\label{fig:mass_loss_rate_1D}
\end{figure}

\Cref{fig:mass_loss_rate_1D} shows both contributions to the mass loss rate. For larger density ratios $\mu$, the first term in \cref{eq:m_loss_2} (dashed colored lines) dominates. With decreasing $\mu$, the second term (shaded region) becomes more important. The mass loss rate due to mass transfer through the tidal radius is close to the prediction in \cite{Hui:2016ltb} which they attribute to tunneling effects, as long as we take the decreasing of the core's central density into account. It is only about $30\%$ larger than the prediction. \footnote{The results were confirmed by solving the Poisson equation using a Numerov Algorithm, while the wave function was evolved with a fully implicit Crank-Nicholson scheme. The same imaginary potential \cref{eq:boundary_con} was used.}

\subsection{Full three-dimensional case}
\label{sec:3D}

We now consider the three-dimensional case (full model). There are two major differences between the simplified model and the full model: (1) in the former, the subhalo is assumed to be in a state of synchronous rotation, i.e. the subhalo rotates like a rigid body ($\mathbf{\nabla}\times\mathbf{v}\neq 0$), which cannot be true for FDM cores whose velocity field is curl-free; (2) the simplified model assumes a spherically symmetric tidal force, so the solitonic core spins at a constant rate.
On the contrary, in the full model, the solitonic core can spin up due to tidal torque (see Sec. \ref{sec:spinning} for detail).  

\begin{figure}[htbp]
\includegraphics[width=\columnwidth]{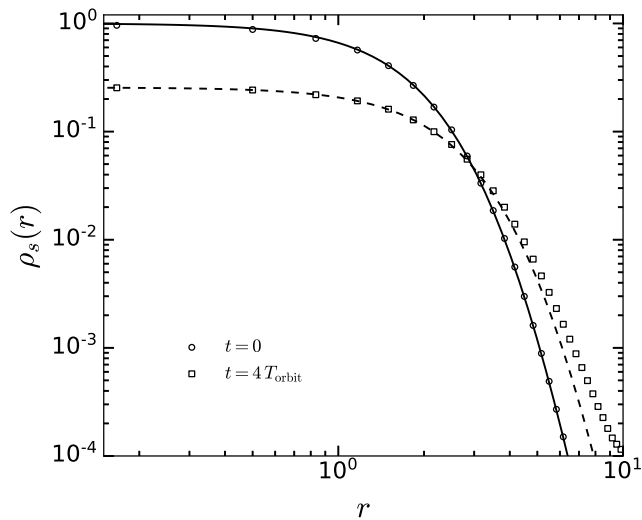}
\caption{The density profile of the core at the initial time and at $t=4\,T_{\rm orbit}$ for a density ratio $\mu=50$. The circles and squares show the average radial density profile obtained from the simulation. The lines display fitted profiles defined in \cref{eq:rho_c}. It can be seen that the cores are well described by soliton profiles even after losing substantial amounts of mass.}
\label{fig:rho_r_50}
\end{figure}

In the three-dimensional case, it is difficult to find a well-defined tidal radius. Therefore, instead of analyzing the mass within the tidal radius, we will look at the evolution of the core mass $M_c$, defined as the enclosed mass within the core radius $r_c$ (about $1/4$ of the total soliton mass). In our simulations, we find that after the core loses some of its mass, it quickly relaxes to a new soliton profile with a smaller central density (see \cref{fig:rho_r_50}).
From the density profile of solitons, \cref{eq:rho_c}, we can see that the core mass $M_c\propto\rho_c^{1/4}$. Therefore we need to adjust \cref{eq:m_loss} accordingly:
\begin{equation}
\frac{\dot{M}_{c}}{M_{c}}=\frac{1}{4}\frac{\dot{\rho}_{c}}{\rho_{c}}=\frac{1}{2}\,{\rm Im}(E)\,.
\label{eq:m_loss2}
\end{equation}

\begin{figure}[htbp]
\includegraphics[width=\columnwidth]{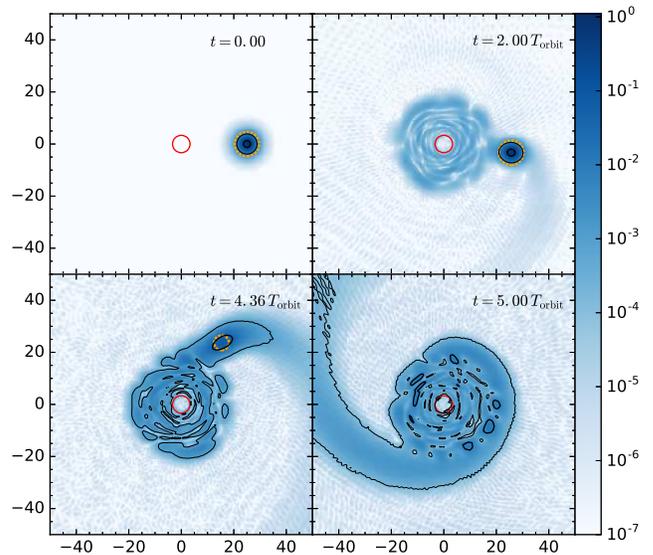}
\caption{Slices through the density field at different times for an initial density ratio $\mu=50$. The circle in the center of each plot indicates the size of the host (for simplicity the host is treated as a small sphere with uniform density). The thick and thin contour lines mark where the density drops to $50\%$ (core radius) and $1\%$ of the maximum density, respectively. The dashed circles show the tidal radii computed from the spherically symmetric approximation.}
\label{fig:evo_50}
\end{figure}

\Cref{fig:evo_50} presents slices through the density field at different times for $\mu=50$. The thick and thin contour lines mark where the density drops to $50\%$ (core radius) and $1\%$ of the maximum density, respectively. For comparison, we also show the tidal radius computed from the spherically symmetric approximation (dashed circles). As can be seen, the core loses mass gradually but since the gravitational time scale is smaller than the mass loss time scale, it quickly relaxes to a new soliton state with a lower central density (upper-right plot, see also \cref{fig:rho_r_50}). At $t=4.36\,T_{\rm orbit}$ (lower-left plot), the tidal radius is comparable to the core radius. Afterwards, in less than one orbit, the core is totally disrupted and leaves only a long tail behind (lower-right plot).

\begin{figure}[htbp]
\includegraphics[width=\columnwidth]{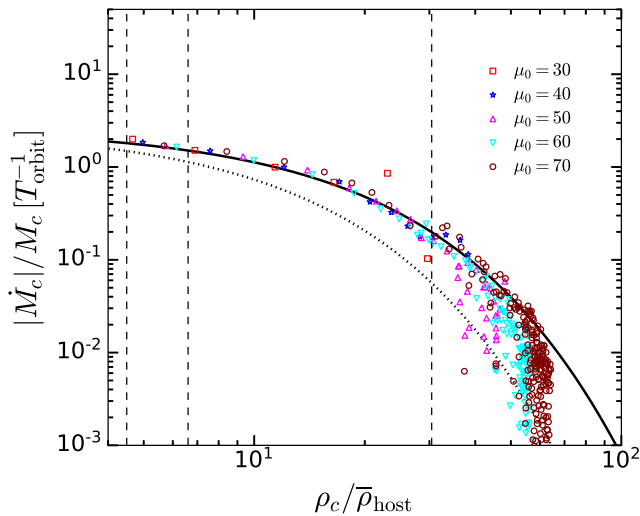}
\caption{Core mass loss rate for different initial density ratios. The lines show the prediction from \cref{eq:m_loss_fitting} and \cref{eq:m_loss2} with $\gamma=3/2$ (solid line) and $\gamma=1$ (dashed line). The vertical dashed lines are the same as in \cref{fig:pic_r_t}.}
\label{fig:mass_loss_rate}
\end{figure}

\Cref{fig:mass_loss_rate} illustrates the core mass loss rate from simulations with different initial conditions compared to the prediction from \cref{eq:m_loss_fitting,eq:m_loss2}. In general, the results are close to the predictions.
At very early times, the core mass decreases more slowly than the prediction. This can be attributed to the initial conditions. We initially assume a soliton without self-rotation, i.e. the proportionality coefficient in the tidal potential should be $\gamma = 1$ at the beginning. The core subsequently acquires angular momentum and starts to spin up due to tidal torque, so $\gamma$ approaches ${3}/{2}$. For comparison, the dotted line shows the prediction with $\gamma=1$. As can be seen, at early times the mass loss rate computed from simulations roughly falls between the solid and dotted curves.

\section{Tidal locking}
\label{sec:spinning}

\begin{figure}[htbp]
\includegraphics[width=\columnwidth]{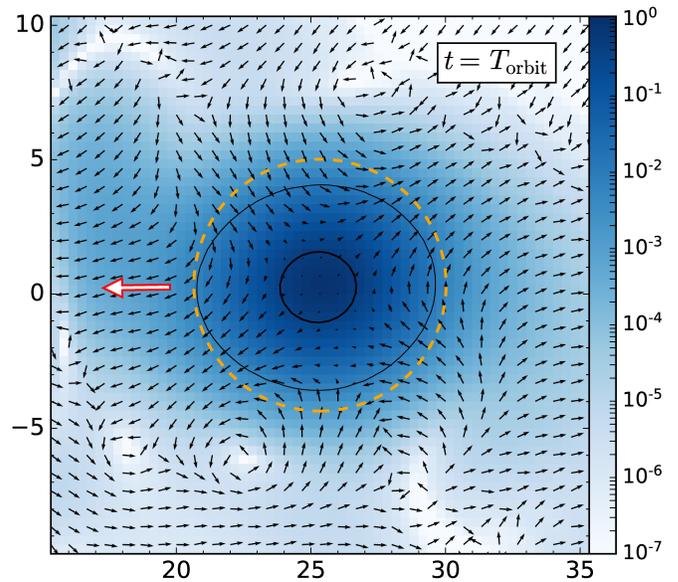}
\caption{Slice through the core. The color map indicates the density while the black arrows trace the velocity field relative to the core's collective motion. Inside the tidal radius (dashed circle) the velocity field is characteristic for an irrotational Riemann-S ellipsoid \cite{RindlerDaller:2011kx}. Outside the core, vortices can be seen. The thick black circle in the middle represents the core radius while the thin black ellipsoid marks the area where the density drops to one percent of the central density, i.e. almost all the mass lies within the tidal radius. The hollow arrow points towards the host's center.}
\label{fig:velocity}
\end{figure}

One important difference between FDM subhalo cores and rigid-body satellites is that the solitonic core does not sustain uniform self-rotation. In our simulations, we find that an initially spherical solitonic core without self-rotation gradually spins up and forms an irrotational ellipsoid in the tidal field of the host. The velocity field in a typical simulation with $\mu=50$ can be seen in \cref{fig:velocity}.
Inside the tidal radius (dashed circle, computed from the spherically symmetric approximation), the velocity field is characteristic for an irrotational Riemann-S ellipsoid \cite{RindlerDaller:2011kx}. 
The core is elongated towards the host's center (the hollow arrow in \cref{fig:velocity}), indicating that the core is tidally locked. However, unlike a rigid body, the core does not rotate uniformly.

\begin{figure}[htbp]
\includegraphics[width=\columnwidth]{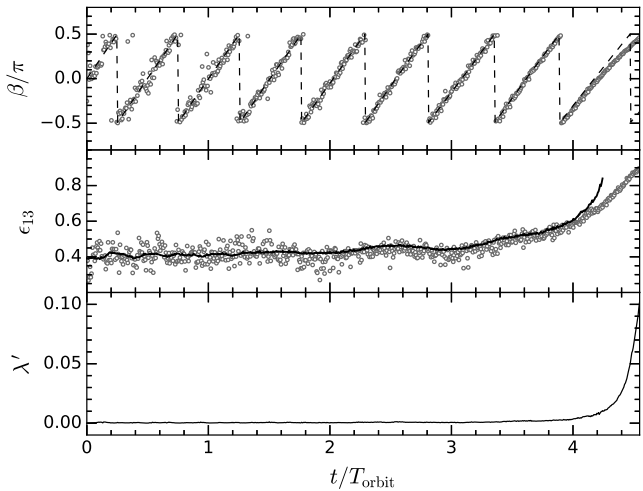}
\caption{Representative run with initial density ratio $\mu=50$. Top: angle between the longest principal axis of the core and $x$-axis (grey circles). The dashed line shows the angle between the line joining the center of the core and the center of host and $x$-axis. It can be seen that the core is tidally locked most of the time. Center: eccentricity $\epsilon_{13}$ of the core. The solid line shows the expected values from \cref{eq:ellip_roche}. Bottom: spin parameter of the core.}
\label{fig:spin}
\end{figure}

As the central density of the core decreases over time, the shape of the ellipsoid also changes. Denote the semiaxes of the core as $a_1$, $a_2$, and $a_3$ ($a_1\geq a_2\geq a_3$). Then the change can be characterized by the eccentricities of the ellipsoid. In the middle panel of \cref{fig:spin}, we show the evolution of the eccentricity of the ellipsoid in the plane that contains the shortest and longest principal axes, i.e. $\epsilon_{13}=[1-(a_3/a_1)^2]^{1/2}$. We find that the eccentricity $\epsilon_{12}=[1-(a_2/a_1)^2]^{1/2}$ is very close to $\epsilon_{13}$, implying that the core is approximately spheroidal.

Assuming constant densities $\rho_{h}$ and $\rho_{\rm sat}$ for the host and satellite, Roche found that the ellipticity of an equilibrated, tidally locked, fluid satellite can be calculated analytically as a function of its density ratio \cite{Roche:1850}
\begin{equation}
    \rho_{h}/\rho_{\rm sat} = \frac{1 - \epsilon^2}{2\epsilon^3} \left[ \left(3-\epsilon^2 \right) \operatorname{artanh} \epsilon -3 \epsilon \right]\,.
    \label{eq:ellip_roche}
\end{equation}
Since the satellite's density inside the core radius does not change significantly, we set $\rho_{h}/\rho_{\rm sat}=\mu$ and calculate the expected ellipticity from \cref{eq:ellip_roche}. The center panel of \cref{fig:spin} confirms that this approximate solution agrees well with our numerical results as long as the core stays tidally locked.

The rotation of the core can be parameterized by the dimensionless spin parameter as defined in \cite{Bullock:2000ry} for DM halos
\begin{equation}
\lambda' = \frac{L_c}{\sqrt{2}M_c V R}\,,
\label{eq:lambda}
\end{equation}
where $L_c$ is the core angular momentum with respect to its center, $R\equiv(a_1 a_2 a_3)^{1/3}$ is the mean core radius, and $V$ is the circular velocity at $R$. We show the evolution of $\lambda'$ in the lower panel of \cref{fig:spin}.

The top panel of \cref{fig:spin} displays the angle between the longest principal axis of the core and $x$-axis, compared to the angle between the line joining the center of the core and the
$x$-axis. It can be seen that the core becomes tidally locked in less than $1/4\,T_{\rm orbit}$. 

From the center panel of \cref{fig:spin}, we see that the core eccentricity increases over time. In order for the core to stay tidally locked, the spin parameter has to increase as well. Thus the core will slightly deviate from tidal locking until it obtains additional angular momentum due to tidal torque and becomes tidally locked again. At late time, angular momentum transfer from orbital motion to self-rotation of the satellite becomes insufficient to maintain tidal locking. The core rotation lags behind its orbital frequency and the core quickly becomes tidally disrupted. 

\section{Satellite galaxies in the Milky Way}\label{sec:sat_galaxy}

Having quantified the core mass loss rate solely depending on the ratio $\mu$ between the central density $\rho_{c}$ of the satellite's core and the average host density $\overline{\rho}_{\rm host}$, it is straightforward to estimate the survival time of satellite galaxies in the Milky Way. Assuming a host mass of $M_{\rm host}$, a given satellite distance $D$ to the galactic center directly translates into a mean host density $\overline{\rho}_{\rm host}$ within the satellite's orbit. If we further assume that the satellite's core evolves along the fitting curve in \cref{fig:mass_loss_rate}, we can compute the minimum central density of the satellite required to survive for $N_{\rm sur}$ orbits:
\begin{equation}
\rho_{c,{\rm min}}=\mu_{\rm min}(N_{\rm sur})\overline{\rho}_{\rm host}=\mu_{\rm min}(N_{\rm sur})\frac{3 M_{\rm host}}{4\pi D^3}\,,
\label{eq:rho_c_min}
\end{equation}
where $\mu_{\rm min}(N_{\rm sur})$ is the minimum density ratio required. Furthermore, for a fixed FDM particle mass $m_{22}\equiv m/(10^{-22}{\rm eV})$, $\rho_{c}$ determines the core mass $M_c\propto\rho_c^{1/4}$ \cite{Schive2014a}. Thus, the mass of the core surviving for $N_{\rm sur}$ orbits must satisfy
\begin{align}
    M_c &>  5.82\times10^8 \left[\mu_{\rm min}(N_{\rm sur})\right]^{1/4} m_{22}^{-3/2} \left(\frac{D}{\rm kpc}\right)^{-3/4}\nonumber\\
        &\left(\frac{M_{\rm host}}{10^{12} M_{\odot}}\right)^{1/4} M_{\odot}\,.
\label{eq:M_c_min}
\end{align}
If we consider the satellite to be disrupted when its core loses $90\%$ of its mass and take $\gamma=3/2$ and $N_{\rm sur}=10$, we find $\mu_{\rm min}=74$ which is slightly larger than estimated in \cite{Hui:2016ltb}. Taking $\gamma=1$ and $N_{\rm sur}=1$, we get a more conservative constraint $\mu_{\rm min}=8.4$.

\begin{figure}[htbp]
\includegraphics[width=\columnwidth]{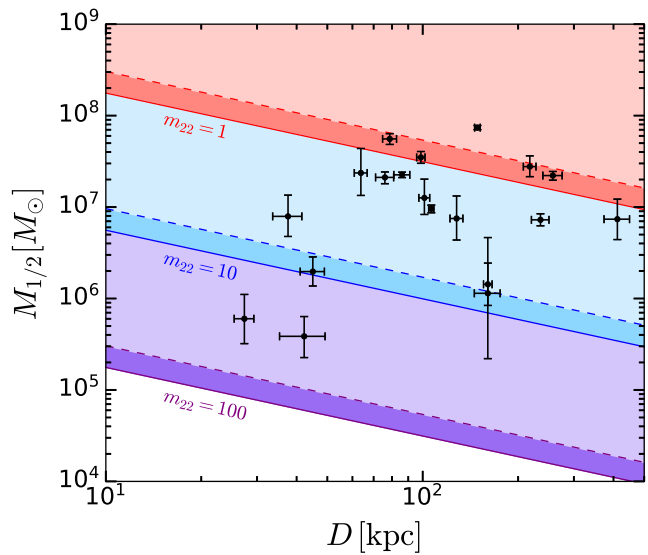}
\caption{Minimum mass of cores that can survive for $N_{\rm sur}$ orbits assuming different FDM particle masses $m_{22}\equiv m/(10^{-22}{\rm eV})$ versus the distance to the Galactic center $D$. For comparison, we also show the half-light mass $M_{1/2}$ of some satellite galaxies in the Milky Way \cite{Wolf:2010}. The mass of the host is taken to be $10^{12} M_{\odot}$. For each particle mass, the solid curve is obtained by assuming $\gamma=1$ and $N_{\rm sur}=1$ while the dashed curve is obtained by assuming $\gamma=3/2$ and $N_{\rm sur}=10$.}
\label{fig:Mc_min}
\end{figure}

In \cref{fig:Mc_min}, we use \cref{eq:M_c_min} for different FDM particle masses to constrain the minimum mass of cores that can survive for $N_{\rm sur}$ orbits as a function of the distance to the Galactic center. The corresponding satellite mass should be larger than its core mass. We consider a Milky Way--like host $M_{\rm host}=10^{12}M_{\odot}$. For comparison, we also show the half-light mass $M_{1/2}$ of some satellite galaxies in the Milky Way \cite{Wolf:2010}. As expected, satellites close to the Galactic center are particularly susceptible to tidal disruption and therefore place the most stringent constrains on the particle mass. Specifically, the lightest satellites close to the Galactic center will only survive for more than one orbital time if the particle is as heavy as $m\simeq 2\times 10^{-21}$ eV.

With knowledge of the distribution of the initial mass of subhalos accreted by a Milky Way--like host, it will be possible to predict the probability that we can find a satellite galaxy with a given mass in the Milky Way depending on the dark matter particle mass. This can be done either by performing simulations like ours with appropriate parameters for initial conditions \cite{Shaun:2017}, or by implementing the mass loss rate found in this work in semianalytic models (SAMs) and computing the SHMF \cite{Du:2016zcv}.

\begin{figure}[htbp]
\includegraphics[width=\columnwidth]{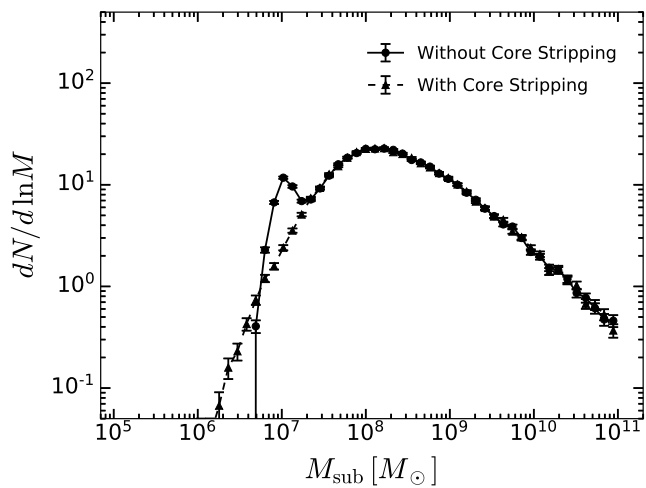}
\caption{Subhalo mass function for $m_{22}=10$ with (dashed line) and without (solid line) including the tidal stripping of subhalo cores.}
\label{fig:SHMF}
\end{figure}

\Cref{fig:SHMF} gives an example of how the tidal stripping of cores affects the SHMF for $m_{22}=10$. The solid line is obtained from SAMs assuming the cores of subhalos are stable against tidal stripping as in \cite{Du:2016zcv}. As can be seen, the SHMF exhibits a peak at about $M_{\rm sub}=10^7 M_{\odot}$ corresponding to subhalos that consist only of their stable cores. Instead, if we include the tidal mass loss of subhalo cores, the peak of the SHMF at lower masses is smeared out while the SHMF is not affected at higher masses (dashed line). Here we have assumed a mass lose rate given by \cref{eq:m_loss_fitting,eq:m_loss2} with $\gamma=3/2$ which 
is a good approximation to the core mass loss rate (see \cref{fig:mass_loss_rate}). More detailed analysis of the SHMF and possible constraints from observations will be discussed in a forthcoming paper \cite{du:2018}.

\section{Conclusions}
\label{sec:conclusions}

We investigated the tidal disruption of fuzzy dark matter (FDM) subhalo cores numerically using a fourth-order pseudospectral method.
First, we considered an idealized case with a spherical tidal potential. We calculated the mass loss of the core resulting from mass transfer through the tidal radius and decreasing tidal radius, and found that the contribution from the former is close to the semianalytic prediction of \cite{Hui:2016ltb} if the decreasing density ratio is taken into account.
For lower density ratios, the mass loss due to a decreasing tidal radius dominates. In general, the core loses mass more quickly than estimated in \cite{Hui:2016ltb} since while the core loses mass, its central density decreases, making the core more vulnerable to tidal forces (see \cref{fig:rho_c_t_Phi}).

We also performed three-dimensional simulations of a more realistic case where the core is evolved in the central potential of a host, treated as a small uniform sphere. In this case, it is difficult to find a well-defined tidal radius contrary to the previous case with spherical symmetry.
The results show that when the solitonic core loses mass, it rapidly equilibrates to a new solitonic state with lower central density (\cref{fig:pic_r_t}). Even after losing a substantial fraction of its initial mass, the average core density profile can still be fitted by a solitonic profile. Therefore, instead of computing the mass loss rate of the matter within the tidal radius, we calculated the evolution of the core mass $M_c\propto\rho_c^{1/4}$ with $\rho_c$ the central density of the core. The mass loss rate as a function of the density ratio $\mu$ can be well described by the semianalytic prediction of \cite{Hui:2016ltb} if we account for a dynamically varying density ratio and an extra factor of $1/4$ coming from the scaling relations of solitonic cores (see \cref{fig:mass_loss_rate}).

Additionally, \cite{Hui:2016ltb} assume a spherical tidal potential $\Phi_t=-\gamma\omega^2 r^2$ with $\gamma=3/2$. They thus model the satellite as a rigid body that orbits the host with synchronous rotation
which cannot be satisfied by FDM cores whose velocity field obeys $\mathbf{\nabla}\times\mathbf{v}=0$. For a satellite without self-rotation, we have $\gamma=1$. Thus in general, $\gamma$ varies between $1$ and $3/2$ depending on the internal velocity of the core.

Finally, we found that initially non-rotating cores acquire angular momentum in a tidal field due to tidal torque. The cores become tidally locked in less than $1/4\,T_{\rm orbit}$. The internal velocity field is described by an irrotational Riemann-S ellipsoid instead of a uniformly rotating rigid body satellite. With decreasing central density, the eccentricity of the ellipsoid increases and can be well approximated by a Roche ellipsoid as long as the core is approximately tidally locked. At later times, the core cannot gain sufficient additional angular momentum and begins to deviate from tidal locking. It is then rapidly disrupted.

In the three-dimensional simulations, we assumed the host to be a small uniform sphere. This is a good approximation as long as the subhalo is not too close to the center of the host. We also tested an NFW potential for the host. In this case, the tidal force is slightly smaller due to the non-vanishing density of the host at the position of the satellite. This difference can be accounted for by a redefinition of the density ratio $\mu_{\rm eff}\equiv\rho_c/\rho_{\rm eff}=\rho_c/[\overline{\rho}_{\rm host}-\rho_{\rm host}(r_{\rm sat})]$ and an effective orbital angular velocity $\omega_{\rm eff}=({4}/{3}\,\pi G \rho_{\rm eff})^{1/2}$. With these redefinitions, the results are consistent with the approximation of the host as a uniform sphere.

Our results can be used to estimate a lower bound on the mass of satellite galaxies that can be observed in the host galaxy in FDM scenarios. We calculated the minimum mass of cores that can survive for a given number of orbits in a Milky Way--like host. Its value depends on the FDM particle mass and the distance to the center of the host. We compare it with observed satellite galaxies in the Milky Way (see \cref{fig:Mc_min}). Our results are useful for finding constraints on FDM from the observational abundance of satellite galaxies in the Milky Way.

\acknowledgements
We thank Shaun Hotchkiss and Jan Veltmaat for helpful discussions. X.D. thanks Jiajun Chen for useful discussions on code development. X.D. acknowledges the China Scholarship Council (CSC) for financial support.

\appendix

\section{CONVERGENCE TEST}
\label{sec:appendix}

In this work, we solve the Schr\"{o}dinger-Poisson equations with a fourth-order pseudospectral method as described in Sec. \ref{sec:sim}. To test the convergence of our code in time, we simulate mergers of multiple solitons, which have been studied in detail in previous papers, e.g. \cite{Schive2014b,Schwabe:2016rze,Mocz:2017wlg}. The simulated box has a length of $40$ on each side and a resolution of $240^3$ cells. At the initial time, $20$ solitons with the same core radius $r_c=1.308$ are randomly put in the box. We check the conservation of the total energy and compare it with the well-known second-order algorithm, kick-drift-kick leapfrog method.

\begin{figure}[htbp]
\includegraphics[width=\columnwidth]{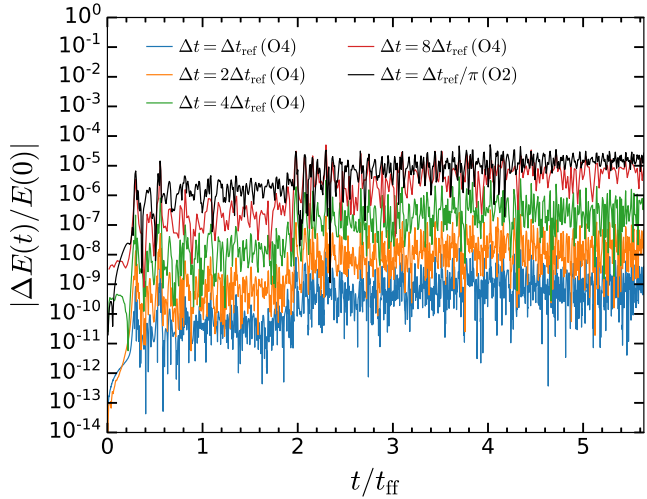}
\caption{Numerical error of the total energy with respect to time for different time step sizes and different algorithms. The time is in units of the free-fall time scale $t_{\rm ff}=\sqrt{3\pi/(32 G \rho)}$ with $\rho$ equal to the average density over the whole simulated box. Here ``O4" refers to the fourth-order algorithm we used in our simulations (Sec.\ref{eq:exp_H_M}). ``O2" refers to the second-order kick-drift-kick formulation \cref{eq:exp_H_O2}, which is widely used in previous simulations.}
\label{fig:error_t}
\end{figure}

\Cref{fig:error_t} shows the relative error of the total energy with respect to time for the fourth-order algorithm (O4, colored lines) and the second-order algorithm (O2, black line). Results from simulations with different time step sizes are shown. As can be seen, the fourth-order algorithm has comparable accuracy to the second-order algorithm even if the time step size is $8\pi$ times larger.

\begin{figure}[htbp]
\includegraphics[width=\columnwidth]{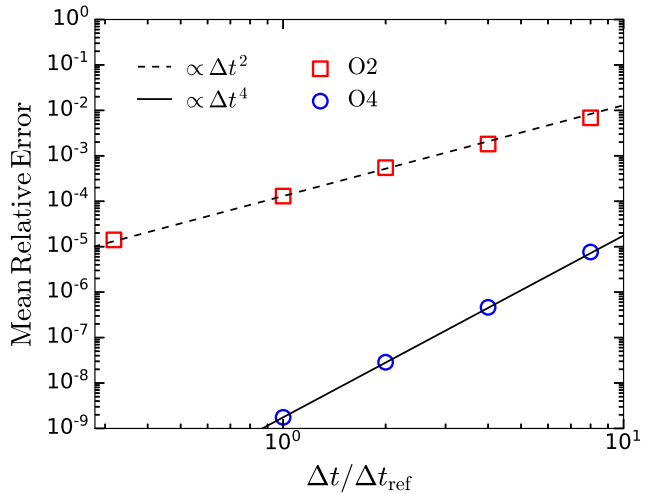}
\caption{Average numerical error of the total energy with respect to the time step size. Only data with $t>3\,t_{\rm ff}$ when the numerical error oscillates around roughly a constant value is included in the analysis.}
\label{fig:error_tend}
\end{figure}

\Cref{fig:error_tend} shows the average relative error of the total energy with respect to the time step size. As expected the algorithm we implemented (circles) has fourth-order convergence. Compared to the second-order algorithm (squares), it converges faster and the error is several orders of magnitude smaller if the same time step size is used.

\clearpage
\bibliography{tidal_disruption}

\end{document}